\documentstyle{article}
\begin{document}

\begin{titlepage}
\title{Electron capture in $GaAs$ quantum wells via electron-electron and optic
phonon scattering}
\author{K. K\'alna and M. Mo\v sko \\Institute of Electrical Engineering,
Slovak Academy of Sciences\\ D\'ubravsk\'a cesta 9, Sk-842 39 Bratislava,
Slovakia, \and F. M. Peeters \\ Department of Physics, University of Antwerp
(UIA),\\ Universiteitsplein 1, B-2610 Antwerpen-Wilrijk, Belgium}
\maketitle
\begin{abstract}
\noindent Electron capture times in a separate confinement quantum well (QW)
structure with finite electron density are calculated for electron-electron
(e-e) and electron-polar optic phonon (e-pop) scattering. We find that the
capture time oscillates as function of the QW width for both processes with the
same period, but with very different amplitudes. For an electron density of
$10^{11} cm^{-2}$ the e-e capture time is $10^1-10^3$ times larger than the
e-pop capture time except for QW widths near the resonance minima, where it is
only $2-3$ times larger. With increasing electron density the e-e capture time
decreases and near the resonance becomes smaller than the e-pop capture time.
Our e-e capture time values are two-to-three orders of magnitude larger than
previous results of Blom~{\it et~al.} [Appl. Phys. Lett. {\bf 62}, 1490
(1993)].
The role of the e-e capture in QW lasers is therefore readdressed.
\end{abstract}
\end{titlepage}

\newcommand{\be}{\begin{equation}}
\newcommand{\ee}{\end{equation}}
\newcommand{\bd}{\begin{displaymath}}
\newcommand{\ed}{\end{displaymath}}
\newcommand{\ba}{\begin{eqnarray}}
\newcommand{\ea}{\end{eqnarray}}

\newcommand{\k}{{\bf k}}

The electron capture in a quantum well plays an important role in optimizing
the
performance of separate confinement heterostructure quantum well (SCHQW) lasers
\cite{zhao}. Quantum calculations \cite{brum:bastard} of polar optic phonon
(pop) emission induced capture in $GaAs$ QW predicted oscillations of the
capture time versus the QW width, which have been observed \cite{blom:smit}.
The
minima of the oscillations provide the optimum well and barrier width for an
optimized capture efficiency, resulting in an improved modulation response and
a
reduced threshold current of the laser. At high electron densities the
electron-electron (e-e) scattering induced capture is expected to play an
important role. Blom {\it et al.} \cite{blom:haverkort} predicted that the e-e
capture time in a $GaAs$ QW with electron density of $10^{11} cm^{-2}$
oscillates with approximately the same amplitude (and the same period) as the
e-pop mediated capture time. Away from the oscillation minima the e-pop capture
was weak and the e-e capture was expected to increase the threshold current in
the SCHQW laser via excess carrier heating in the QW \cite{blom:haverkort}. The
minima of the e-e capture time oscillations were found to be below $1$ ps,
which
is a promising value for efficient capture. This result is surprising, because
subpicosecond times are typical for intrasubband e-e scattering
\cite{mosko,artaki}, rather than for intersubband e-e scattering
\cite{goodnick:lugli}. Further work is also necessary to clarify the density
dependence of the e-e capture.

In this letter the e-e and e-pop scattering induced capture times are
recalculated for the same SCHQW as in the letter by Blom {\it et al.}
\cite{blom:haverkort}. We find that for an electron density of $10^{11}
cm^{-2}$
the e-e capture time is typically $10^1-10^3$ times larger except for QW widths
near the resonance minima, where it is only $2-3$ times larger. For densities
above $\sim 5 \times 10^{11} cm^{-2}$ the resonant e-e capture time is smaller
than the e-pop capture time. The e-e capture is found to be too weak to cause a
significant excess carrier heating which is in contrast to the conclusion of
Ref.~\ref{blom:haverkort}. Compared to the e-pop scattering limited capture,
the
e-e capture decreases the total capture time for an optimized (resonant) QW
width, with a factor of about $2.9-4.2$ at density of $10^{12} cm^{-2}$.

The analyzed SCHQW consists of the $Al_xGa_{1-x}As/GaAs/Al_xGa_{1-x}As$ QW with
500~\AA\ $Al_xGa_{1-x}As$ barriers, embedded between two semiinfinite $AlAs$
layers. The e-e scattering is treated following the approach of
Ref.~\ref{mosko}. When two electrons in subbands $i, j$ with wave vectors $\k$
and $\k_0$ are scattered to subbands $m, n$ with wave vectors $\k'$ and
$\k'_0$,
the e-e scattering rate of an electron with wave vector $\k$ from subband $i$
to
subband $m$ is given by~\cite{mosko}
\be \lambda_{im} (\k) = \frac{1}{N_S A} \sum_{j,n,\k_0} f_j(\k_0)
\lambda_{ijmn}
(g) \qquad, \label{lamee} \ee
where $g=|\k - \k_0|$,
\ba \lambda_{ijmn} (g) = \frac{N_S m^{\ast} e^4}{16\pi \hbar^3 \kappa^2}
\int_0^{2\pi} d\theta\: \frac{F^2_{ijmn} (q)}{q^2 \;\epsilon^2 (q)} \qquad,
\label{paree} \\ q = \frac{1}{2} {\left[ 2g^2 + \frac{4m^{\ast}}{\hbar^2} E_S -
2g { \left(g^2 + \frac{4m^{\ast}}{\hbar^2} E_S \right)}^{1/2} \cos \theta
\right]}^{1/2}, \label{qee} \\
F_{ijmn} (q) = \int^{\infty}_{-\infty} dz\; \int^{\infty}_{-\infty} dz_0\;
\chi_i (z)\:\chi_j (z_0) \;e^{-q|z-z_0|} \;\chi_m (z) \:\chi_n (z_0).
\label{formfac} \ea
$E_S = E_i + E_j - E_m - E_n$, the summation over $\k_0$ includes both spin
orientations, $m^{\ast}$ is the electron effective mass in $GaAs$, $\kappa$ the
static permittivity, $A$ the normalization area, $E_j$ the subband energy and
$f_j(\k_0)$ the electron distribution in subband $j$. Wave functions $\chi_i$
are obtained assuming the $x$-dependent effective mass and flat $\Gamma$-band
with parabolic energy dispersion, both interpolated \cite{hrivnak} between the
$GaAs$ and $AlAs$. To deal with the $0.3$-eV $GaAs$ QW \cite{blom:haverkort} we
take $x=0.305$. The e-e capture time $\tau_{e-e} = \sum_{i,\k} f_i (\k) /
\sum_{i,\k,m} f_i (\k) \lambda_{i,m}(\k) $, where the summation over $i$ ($m$)
includes the subbands above (below) the $AlGaAs$ barrier, and summation over
$j,
n$ in (\ref{lamee}) involves the subbands below the $AlGaAs$ barrier. $f_j
(\k_0)$ is the Fermi function taken at temperature~$8$K and for an electron
density $N_S = 10^{11} cm^{-2}$. $\epsilon (q) = 1+(q_S/q) F_{1111}(q)\;f_1
(\k_0=0)$ is the static screening function due to the electrons in the lowest
subband \cite{mosko}, where $q_S = e^2 m^{\ast}/(2 \pi \kappa \hbar^2)$.

Full circles in Fig.~1 show $\tau_{e-e}$ versus the QW width for $f_i(\k)$
taken
as a constant distribution up to $36.8$ meV above the $AlGaAs$ barrier, which
roughly models the injected "barrier" distribution after a rapid phonon cooling
\cite{brum:bastard,blom:haverkort}. In the inset our calculation is compared
with the result (crosses) of Ref.~\ref{blom:haverkort}. Both $\tau_{e-e}$
curves
oscillate with the QW width and reach a resonant minimum, whenever a new bound
state merges into the QW (the shift of our resonance minima to slightly lower
QW
widths is due to different effective masses in $GaAs$, $AlGaAs$ and $AlAs$,
which we considered when we calculated the electron wave functions). However,
our $\tau_{e-e}$ is two-to-three orders of magnitude larger. The difference of
a
factor of $4$ is due to the missing factor of $1/4$ in the e-e scattering rate
of Ref.~\ref{blom:haverkort} (see Ref.~\ref{mosko} for details). When the
$\tau_{e-e}$ values from Ref.~\ref{blom:haverkort} are multiplied by a factor
of
$4$, our $\tau_{e-e}$ is still $\sim 100$ times larger.

In order to provide insight we consider the QW with width $w=49$ \AA. To
demonstrate how the form factor (see Fig.~2a) affects the e-e scattering rate,
we compare in Fig.~3a $\lambda_{ijmn} (g)$ as obtained using $F^2_{ijmn} (q)$,
shown in Fig.~2a, with $\lambda_{ijmn} (g)$ obtained with $F^2_{ijmn}=1$. The
latter is typically between $\sim 10^{12} s^{-1}$ and $\sim 4\times 10^{12}
s^{-1}$ for all capture transitions and its dependence on $i, j, m, n$ is
simply
manifested through $E_S$. Figure 3a shows a quite different behavior and the
relative importance of the individual capture transitions is determined by the
behavior of the form factors (Fig.~2a). The individual capture times are at
least two orders of magnitude larger than the subpicosecond capture times shown
in Fig.~3b. Subpicosecond e-e scattering is characteristic for intrasubband
transitions as illustrated in Fig.~3 for $\lambda_{1111} (g)$. The form factor
$F_{1111}$ reduces $\lambda_{1111} (g)$ only insignificantly and our
$\lambda_{1111} (g)$ values are close to similar calculations of
Refs.~\ref{artaki} and \ref{moskova}. We believe that Ref.~\ref{blom:haverkort}
predicts much smaller e-e capture times due to a numerical error. It is
straightforward to verify Fig.~3b quantitatively, because formula (\ref{paree})
is reduced to a simple single-integral for $F_{ijmn}=1$. As for the form
factors, calculations are also relatively simple and we can reproduce those
published in Ref.~\ref{blom:haverkort}. It can be seen without calculation that
the results in Fig.~3a have correct order of magnitude, since they naturally
follow from Figs.~3b and 2a.

The e-pop scattering rate of an electron with wave vector $\k$ from subband $i$
to subband $m$ reads \cite{goodnick:lugli,mosk:thesis} (for spontaneous phonon
emission only)
\ba \lambda_{im} (\k) = \frac{e^2 \omega m^{\ast}}{8 \pi \hbar^2} \left(
\frac{1}{\kappa_{\infty}} - \frac{1}{\kappa} \right) \int_0^{2 \pi} d\theta\;
\frac{F_{iimm} (q)}{q \;\epsilon (q) }, \label{lamepop} \\
q = {\left[ 2k^2 + \frac{2m^{\ast}}{\hbar^2} P - 2k { \left(k^2 +
\frac{2m^{\ast}}{\hbar^2} P \right)}^{1/2} \cos \theta \right]}^{1/2},
\label{qepop} \ea
where $P = E_i-E_m-\hbar\omega$,\ $\hbar\omega$ is the pop energy and
$\kappa_{\infty}$ is the high frequency permittivity. We calculate the e-pop
scattering induced capture time $\tau_{e-pop}$ by averaging (\ref{lamepop}) as
discussed for $\tau_{e-e}$. Figure~4 compares $\tau_{e-pop}$ with $\tau_{e-e}$
for the parameters and the constant distribution $f_i (\k)$ from Fig.~1. The
$\tau_{e-pop}$ data shown by empty circles are calculated using the same static
screening $\epsilon (q)$ as for the e-e scattering, empty squares show
$\tau_{e-pop}$ for $\epsilon (q) = 1$. A more accurate calculation with dynamic
screening will give results between these two extreme cases. We conclude that
$\tau_{e-e}$ is one-to-three orders larger than $\tau_{e-pop}$ except for QW
widths near the resonance minima. This conclusion differs from previous
analysis
\cite{blom:haverkort} which predicts nearly the same oscillation amplitude in
both cases. Ref.~\ref{blom:haverkort} also predicts that in the SCHQW lasers
with a QW width below $40$ \AA\ the e-e capture causes significant excess
carrier heating in the QW. Figure~4 does not support this conclusion, because
the e-e capture is negligible.

It is easy to assess the dependence of both capture times on the electron
density $N_S$. For $N_S \geq 10^{11} cm^{-2}$ and temperature $8$ K the static
screening $\epsilon (q)$ is independent on $N_S$, because $f_1 (0) \simeq 1$.
Therefore, the $\tau_{e-pop}$ values in Fig.~4 would be the same also for
higher
$N_S$ and the $\tau_{e-e}$ values would decrease approximately like $N_S^{-1}$
for each QW width. In Fig.~4 we show $\tau_{e-e}$ for $N_S = 2.8 \times 10^{11}
cm^{-2}, 5 \times 10^{11} cm^{-2}$ and $10^{12} cm^{-2}$ only at QW widths of
$43$ \AA\ and $46$ \AA\ in order to save CPU time. At $43$ \AA\ $\tau_{e-e}$ is
much larger than $\tau_{e-pop}$ even for $N_S = 10^{12} cm^{-2}$ due to the
absence of resonance. At $46$ \AA, when the first excited subband merges into
the QW, $\tau_{e-e}$ resonantly decreases about $500$ times and becomes smaller
than $\tau_{e-pop}$ when $N_S \simeq 5 \times 10^{11} cm^{-2}$. When $N_S =
10^{12} cm^{-2}$, the total capture time $\tau_{e-e} \tau_{e-pop} / (\tau_{e-e}
+ \tau_{e-pop})$ is $3.8$~ps for the unscreened e-pop capture ($\tau_{e-pop} =
11$ ps) and $4.3$~ps for the screened e-pop capture ($\tau_{e-pop} = 18$ ps).
Thus, compared to the case $\tau^{-1}_{e-e} = 0$ the capture efficiency of the
QW with the optimized (resonant) width can be improved with a factor $2.9-4.2$
by increasing $N_S$ to $10^{12} cm^{-2}$. At higher densities (not investigated
here) the capture time is expected to increase with $N_S$ on the basis of the
results of Refs.~\ref{sotirelis:hess} and \ref{sotirelis:allmen}. For $N_S >
10^{12} cm^{-2}$ it is no longer justified to treat the e-e and e-pop
scattering
separately \cite{sotirelis:hess,sotirelis:allmen}, because the electrons
interact with a coupled system of electrons and phonons.

The $\tau_{e-pop}$ curve in Fig.~4 does not show a resonant drop for QW widths
$46$~\AA\ and $88$~\AA, because the "barrier" electrons occupy the states below
the threshold for pop emission and cannot be scattered into the subband which
is
in resonance with the top of the QW. A further increase of the QW width shifts
the resonant subband deeper into the QW and the e-pop scattering into this
subband smoothly increases. Resonant decrease of the e-pop capture time takes
place only in the case when the energy distribution of "barrier" electrons is
monoenergetic, with energy slightly lower than the optical phonon energy
\cite{sotirelis:hess}. To show the role of form factors, figure~5 compares the
unscreened e-pop scattering rates obtained using $F_{iimm}$ from Fig.~2b with
the rates obtained with $F_{iimm} = 1$. The individual e-pop capture rates in
Fig.~5a are governed by the relevant form factors, while for $F_{iimm} = 1$
(Fig.~5b) one only finds a simple dependence on $P$. Compared to the e-e
scattering rates in Fig.~3a the corresponding rates in Fig.~5a are
systematically higher, because the e-pop capture rate (\ref{lamepop}) depends
on
$F_{iimm}$ linearly while the e-e scattering rate (\ref{paree}) depends on
$F_{ijmn}$ quadratically. This fact naturally makes the e-e capture less
effective than the e-pop capture except for high electron densities.

In summary, we have compared the e-e and e-pop capture times in the SCHQW. In
both cases the capture time oscillates with the same period, but with very
different amplitude. The e-e capture time is much larger than the e-pop capture
time except for the QW widths near resonances, where it can be even smaller for
electron densities close to $10^{12} cm^{-2}$ which leads to an improved
capture
efficiency of the QW. However, an inefficient e-pop capture in the SCHQW laser
should not lead to excess carrier heating \cite{blom:haverkort} due to e-e
scattering induced capture, because away from the resonance it is still much
stronger than the e-e capture.

Useful discussions with A. Mo\v skov\'a are appreciated. K. K\'alna was
supported by the Open Society Fund, Charta~77 Foundation and Slovak Grant
Agency
for Science. He thanks Professor J. T. Devreese for the kind hospitality during
his stay in Antwerp. M. Mo\v sko is supported by the Slovak Grant Agency for
Science and F. Peeters by the Belgian National Science Foundation.

{\large Figure captions} \vskip1.cm

{\bf Fig. 1.} E-e capture time $\tau_{e-e}$ vs. the QW thickness for $N_S =
10^{11} cm^{-2}$. Full circles show the results for $f_i (\k)$ taken as a
constant distribution up to $36.8$ meV above the barrier. In the inset these
results are compared with the data (crosses) from Ref.~\ref{blom:haverkort}.

{\bf Fig. 2.} (a) Square of the e-e scattering form factor $F_{ijmn} (q)$ as a
function of the wave vector $q$ for a QW with thickness $w=49$~\AA. The indices
$i, j$ and $m, n$ label the initial and final subband states, respectively.
States $1, 2$ are bound in the QW, states $3, 4, \dots, 9$ have subband
energies
above the $AlGaAs$ barrier. Except for the transition $11-11$ all other
transitions are the e-e capture transitions. (b) The e-pop scattering form
factors $F_{iimm} (q)$ [see the text] are shown for comparison.

{\bf Fig. 3.} E-e scattering rate $\lambda_{ijmn}$ vs. the relative wave vector
size $g$. (a) Calculation with form factors from Fig.~2a.\ (b) Calculation with
$F_{ijmn}=1$.

{\bf Fig. 4.} E-pop capture time $\tau_{e-pop}$ and e-e capture time
$\tau_{e-e}$ vs. the QW thickness for $N_S = 10^{11} cm^{-2}$. Open circles
show
$\tau_{e-pop}$ for the statically screened e-pop interaction, open squares show
$\tau_{e-pop}$ for the unscreened e-pop interaction and full circles are the
$\tau_{e-e}$ data from Fig.~1. Crosses, asterisks and pluses at $43$ \AA\ and
$46$ \AA\ show the $\tau_{e-e}$ data for $N_S = 2.8 \times 10^{11} cm^{-2}, 5
\times 10^{11} cm^{-2}$ and $10^{12} cm^{-2}$, respectively.

{\bf Fig. 5.} E-pop scattering rate $\lambda_{im}$ vs. the wave vector $k$ for
the QW with width $w=49$~\AA. (a) Calculation with form factors from Fig.~2b.\
(b) Calculation with $F_{iimm} =1$.

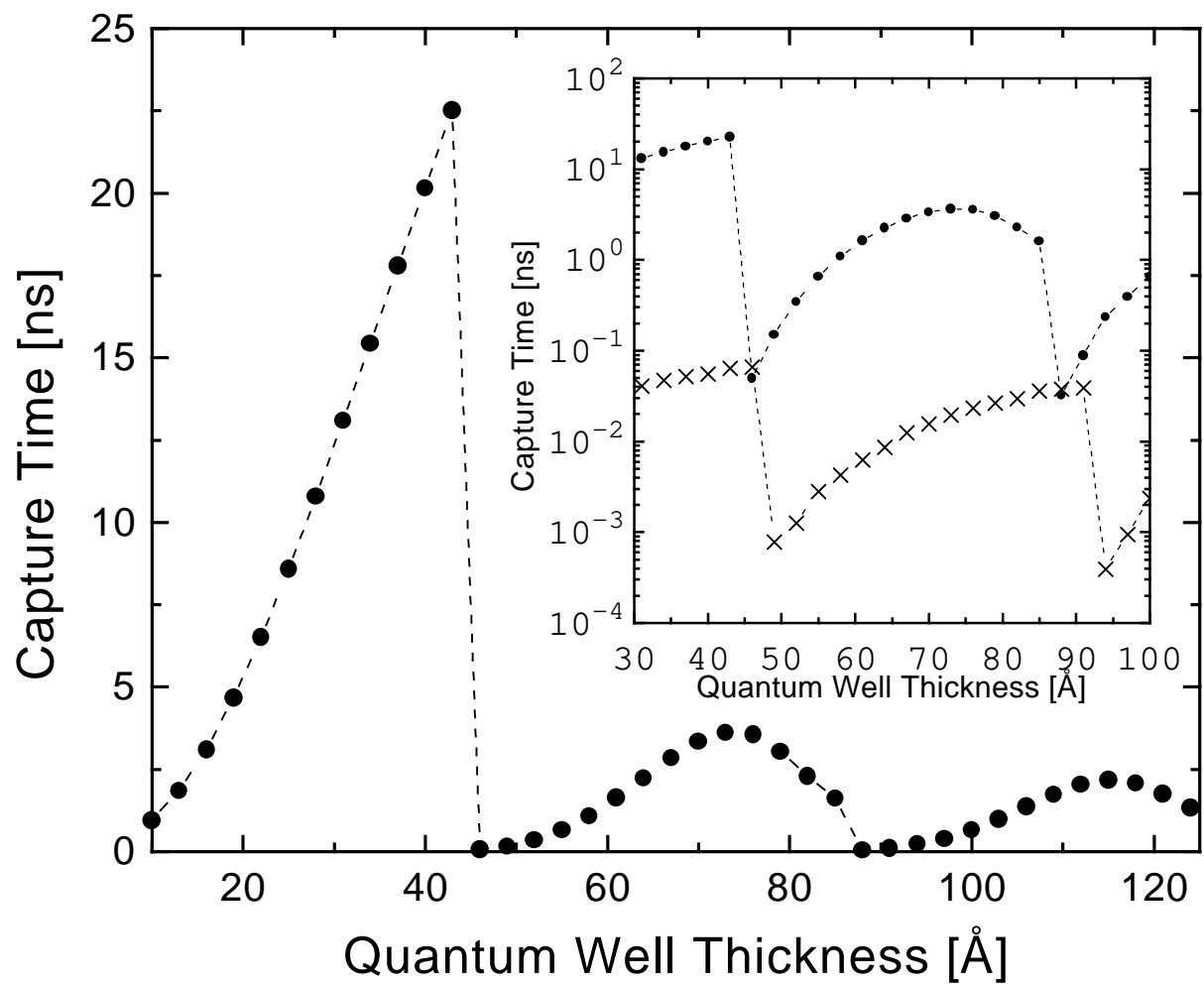
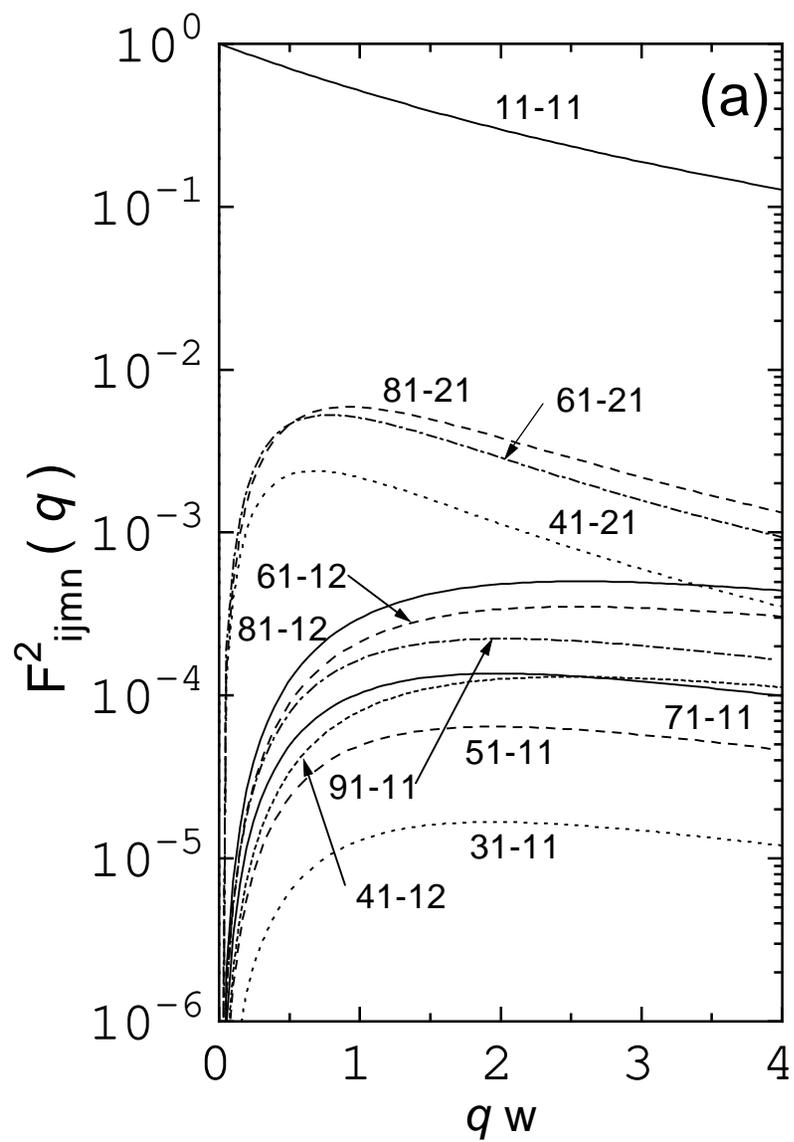
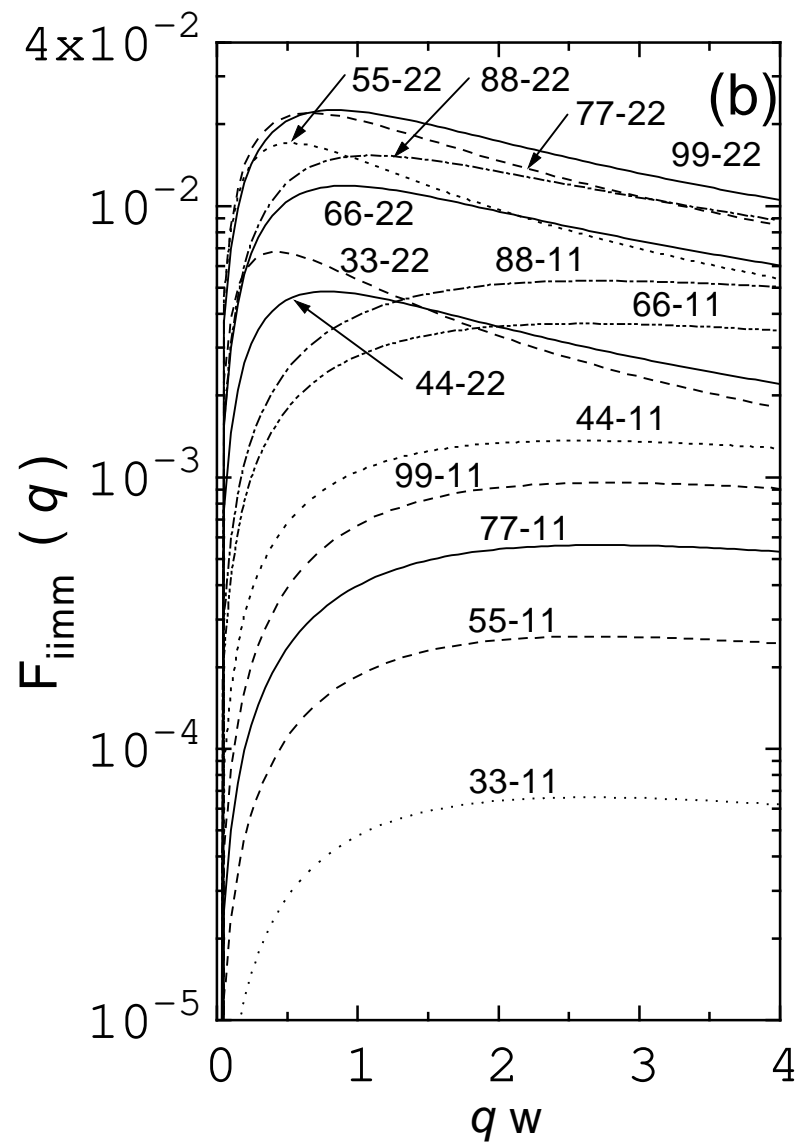
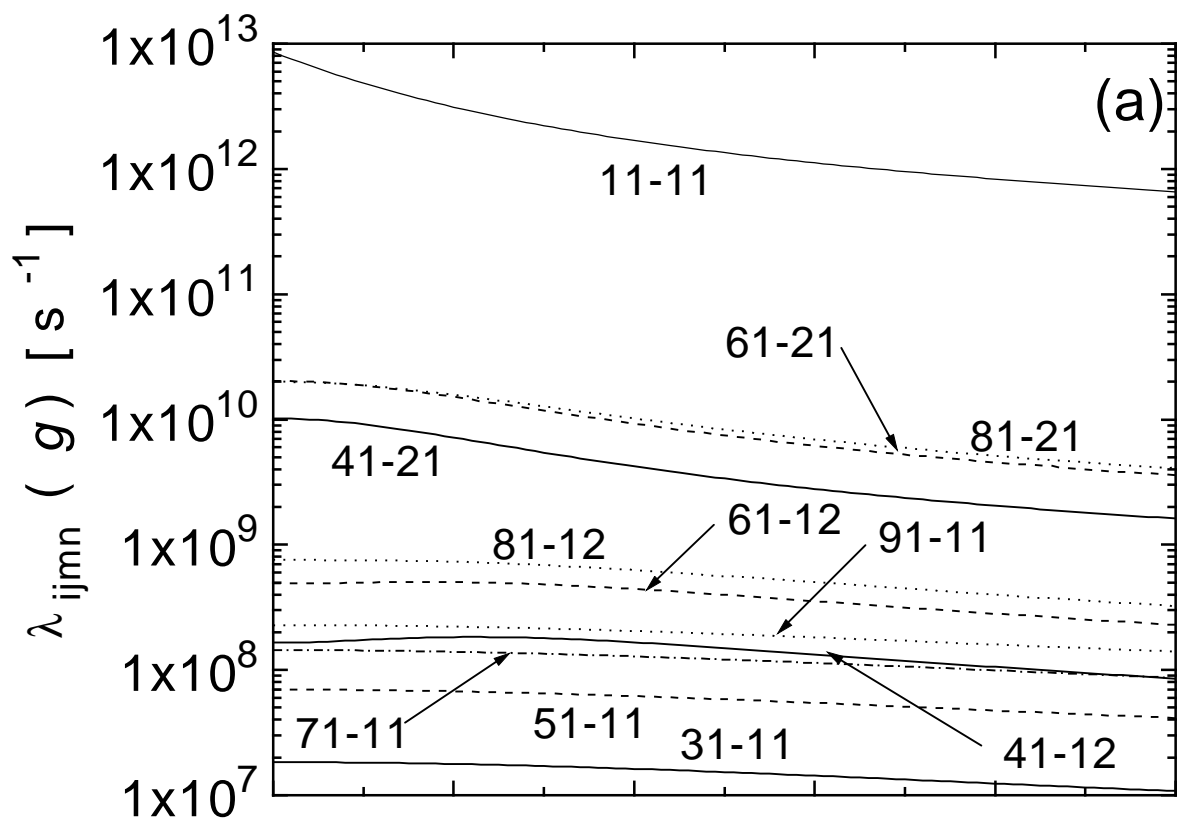
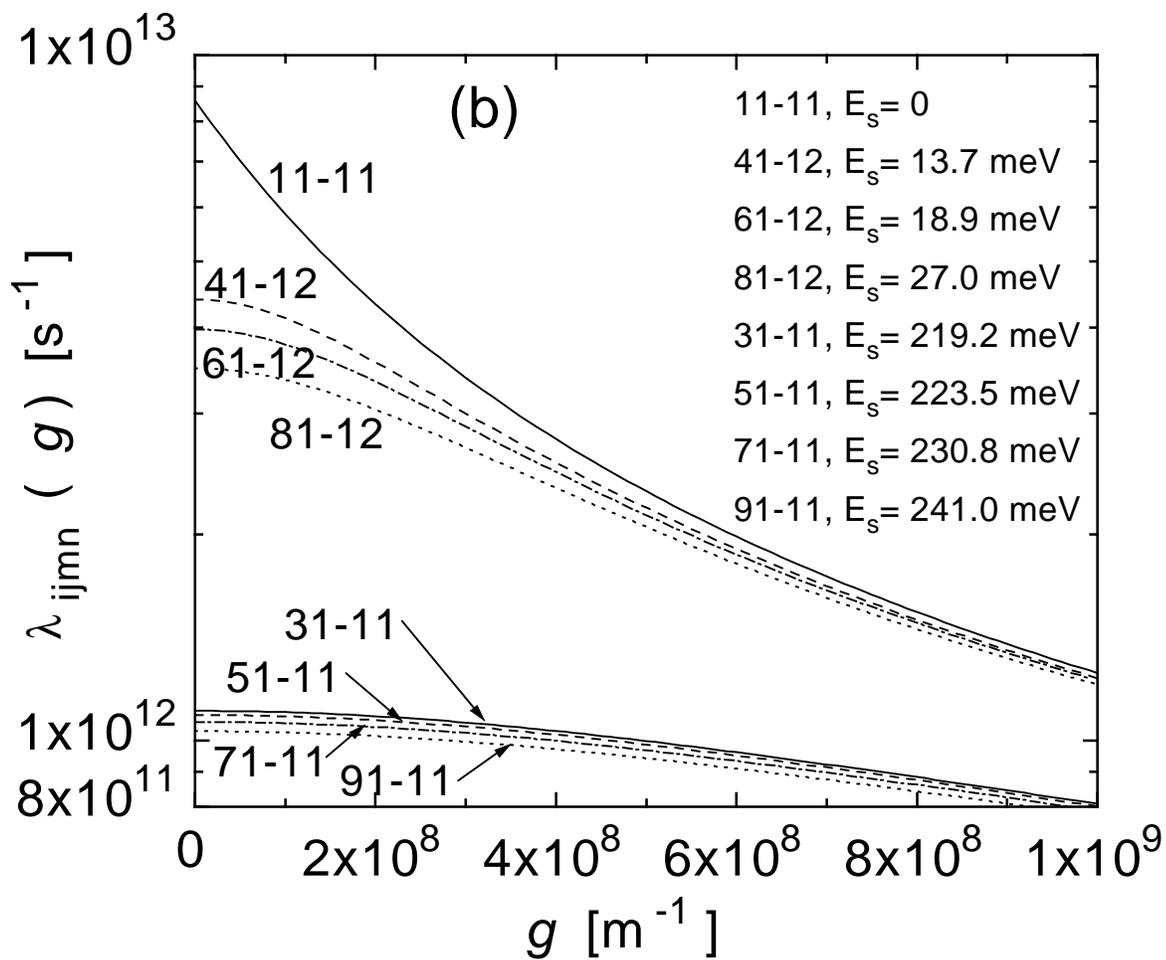
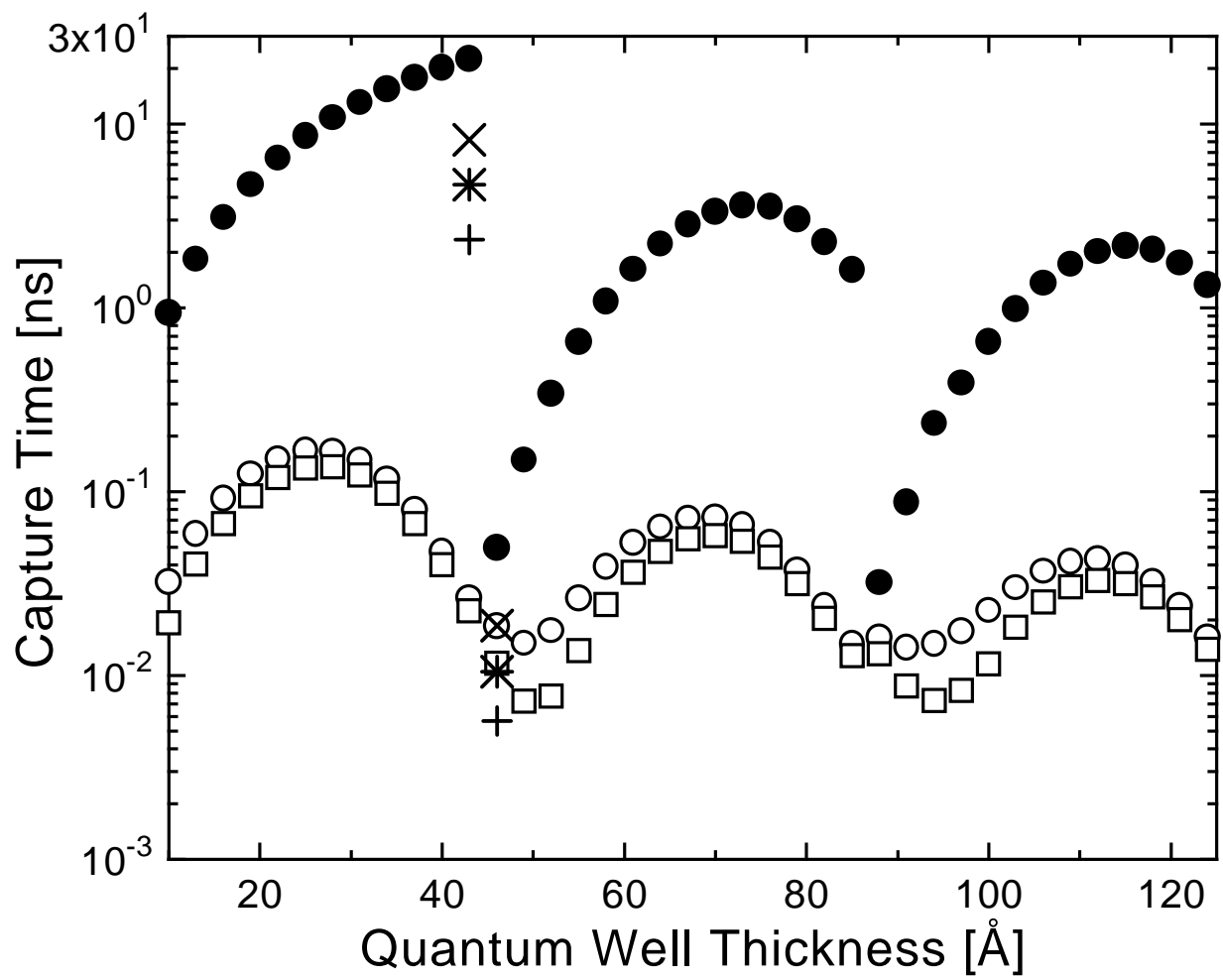
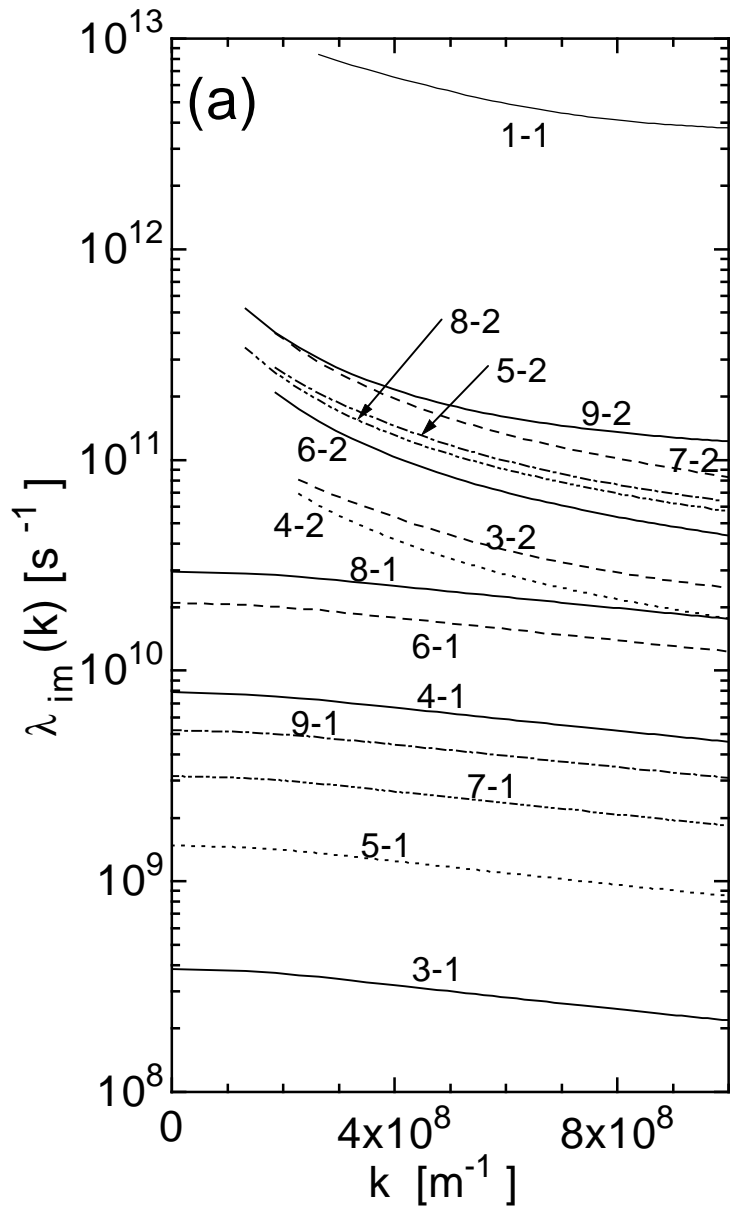
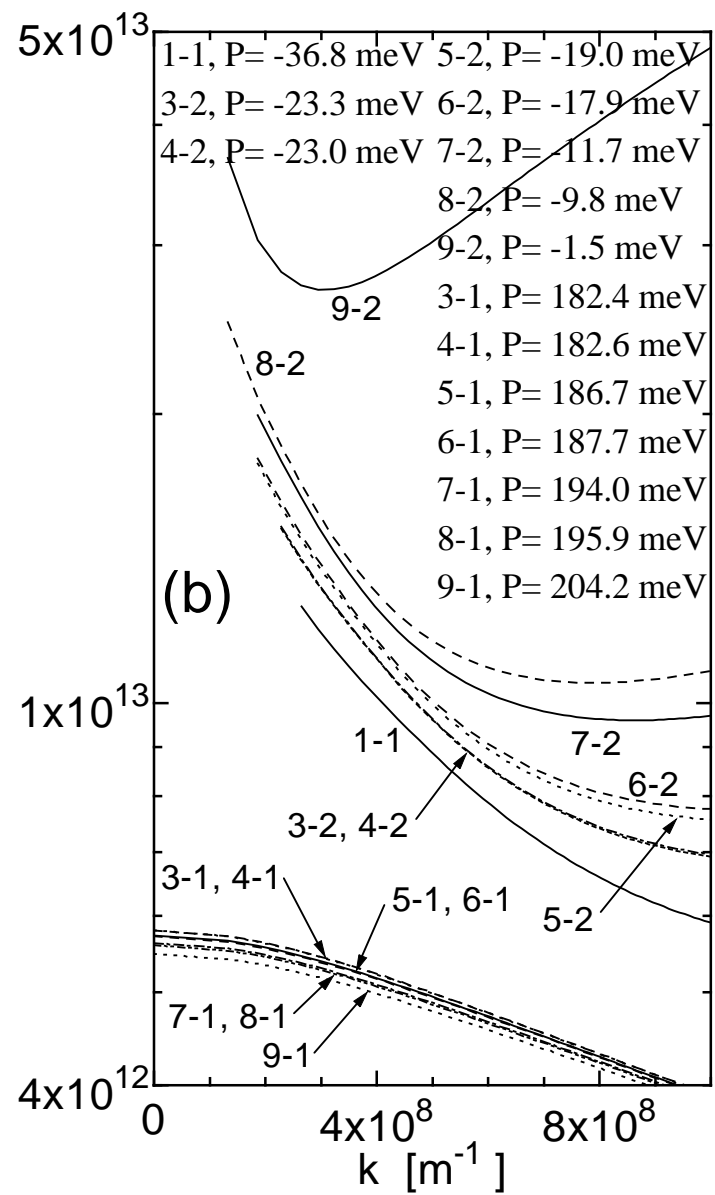
\end{document}